\documentstyle[twocolumn,aps]{revtex}
\begin{document}
\wideabs{
\draft
\title{Mass formulas and thermodynamic treatment
       in the mass-density-dependent model\protect\\
       of strange quark matter}
\author{G. X. Peng,$^{1,2,4}$\ H. C. Chiang,$^{2}$\ J. J. Yang,$^3$\
        L. Li,$^{4}$\ and B. Liu$^{2}$}
\address{
         $^1$China Center of Advanced Science and Technology
               (World Laboratory), Beijing 100080, China \\
         $^2$Institute of High Energy Physics,
               Academia Sinica, Beijing 100039, China \\
         $^3$Department of Physics, Nanjing Normal University,
                                Nanjing 210024, China \\
         $^4$Department of Physics, Nankai University,
                                Tianjin 300071, China \\
         }
%\date{\today}
\maketitle
\begin{abstract}
The previous treatments for strange quark matter in the quark
mass-density-dependent model have unreasonable vacuum limits. We
provide a method to obtain the quark mass parametrizations and
give a self-consistent thermodynamic treatment which includes the
MIT bag model as an extreme. In this treatment, strange quark
matter in bulk still has the possibility of absolute stability.
However, the lower density behavior of the sound velocity is opposite
to previous findings.
\end{abstract}
\pacs{PACS numbers: 24.85.+p, 12.38.Mh, 12.39.Ba, 25.75.-q}
         }

\begin{center}
\section{Introduction}
\end{center}          \label{sec1}

 Since Witten's conjecture that quark matter with strangeness per baryon
of order unity might be bound \cite{Witten}, an extensive body of
literature has investigated the stability and/or probabilities of
strange quark matter (SQM) \cite{PengCCAST}. Because the
application of perturbative quantum chromodynamics (QCD) to
strong-coupling domain is unbelievable while the lattice approach
is presently limited to the case of zero chemical potential, we
have to resort to phenomenological models. One of the most famous
models is the MIT bag model with which Farhi and Jaffe find that SQM
is absolutely stable around the normal nuclear density for a wide
range of parameters \cite{Jaffe}. Further investigations have also
been carried out by many other authors in the bag model
\cite{Madsen,Parija,Schaffner-Bielich}. A recent investigation
indicates a link of SQM to the study of quark condensates
\cite{PengPRC56} while a more recent work has carefully studied
the relation between the charge and critical density of SQM
\cite{PengPRC59}.

  Chakrabarty {\it et al.} \cite{Cha1,Cha2} have
discussed the limitation of the conventional MIT bag model which
assumes that the quarks are asymptotically free within the bag.
In order to incorporate the strong interaction between quarks,
one way is to fall back on the perturbation theory,
 which is questionable in the strong-coupling domain.
An alternative way is to make the quark masses density-dependent.
In this nonperturbative treatment, the strong interaction between
quarks is mimicked by the proper variation of quark masses with
density. There are two questions of crucial importance to this
model. One is how to parametrize quark masses, the other concerns
thermodynamic treatment.  However, the two aspects are not
self-consistent in literature presently.

Here are the popularly used parametrizations for quark masses
 $m_{q}\ (q=u,d,s,)$:
\begin{eqnarray}
 m_{u,d} &=&  \frac{B}{3n_b},  \label{mudcha}     \\
 m_s     &=&  m_{s0}+\frac{B}{3n_b}, \label{mscha}
\end{eqnarray}
where $m_{s0}$ is the $s$ quark current mass, $n_b$ is the baryon
number density, $B$ is the famous MIT bag constant. Equation
(\ref{mudcha}) was first used to study light quark matter
\cite{Fowler}, and later extended to Eq.\ (\ref{mscha}) to
investigate strange quark matter \cite{Cha1,Cha2,Ben}.

 As for the thermodynamic treatment, there exist two controversial
ones in literature up to now.
One %% Chakrabarty {\it et al}.
 expresses the total
pressure of SQM as  \cite{Cha1,Cha2}
\begin{equation}    \label{Pcha}
   P_{\text{1}} = -\Omega,
\end{equation}
where $\Omega$\ is the ordinary thermodynamic potential density of
SQM [see Eq.\ (\ref{Omegai})].
The other adopts the following expression \cite{Ben}:
\begin{equation}    \label{Pben}
   P_{\text{2}} = -\Omega + n_b \frac{\partial\Omega}{\partial n_b}.
\end{equation}
The extra term in Eq.\ (\ref{Pben}) is said to arise from the
baryon density dependence of quark masses. This difference leads
to significantly different results. Therefore, it is meaningful to
take a check of the two thermodynamic treatments.

As is well known, the QCD vacuum is not necessarily empty. To obtain
the vacuum properties, let us take the limit $n_b\rightarrow 0$ for the
two treatments.
It is easy to obtain, at zero temperature, the limits
\begin{eqnarray}
 & \lim\limits_{n_b\rightarrow 0} P_{\text{1}} = 0, &
                                      \label{PchaL} \\
 & \lim\limits_{n_b\rightarrow 0} E_{\text{1}} = B, &
\end{eqnarray}
for the first treatment, and the limits
\begin{eqnarray}
 & \lim\limits_{n_b\rightarrow 0} P_{\text{2}} = -B, &   \\
 & \lim\limits_{n_b\rightarrow 0} E_{\text{2}} = 2B, &
                                        \label{EbenL}
\end{eqnarray}
for the second treatment. Here $E_{\text{1}}$ and $E_{\text{2}}$
are the corresponding energy densities.

According to the fundamental idea of MIT bag model, QCD vacuum has
a constant energy density $B$, the famous bag constant. The mass
parametrization Eq.\ (\ref{mudcha}) is just obtained from this
requirement ($\lim_{n_b\rightarrow 0} E_{\text{1}} \rightarrow
3m_q n_b$ for flavor-symmetric case) \cite{Fowler}. The constant
vacuum energy comes from the fact that QCD vacuum must have a
pressure to maintain pressure balance at the bag boundary.
Obviously, the first treatment can give the correct vacuum energy
and a wrong QCD vacuum pressure. On the contrary, the second
treatment leads to the correct QCD vacuum pressure but a wrong
vacuum energy. In fact, this is just caused by the ignorance of
the QCD vacuum energy which guarantees the pressure balance at the
bag boundary.

It should be pointed out that in getting the unreasonable limits
(\ref{PchaL})--({\ref{EbenL}), we have used the quark mass formulas
(\ref{mudcha}) and (\ref{mscha}). Because these formulas are pure
parametrizations without any real support from underlying
theories, one may ask if the contradictions can be solved by
choosing other parametrizations? According to our present
investigation, one should modify the quark mass formulas and
thermodynamic treatment simultaneously.

  It is the aim of this paper to give a self-consistent treatment
which includes the conventional MIT bag model as an extreme. In
our new treatment, strange quark matter in bulk still has the
possibility of absolute stability. However, the lower density
behavior of the sound velocity in SQM is opposite to previous
findings.

In the following section, we first derive the new quark mass
formulas and describe our thermodynamic treatment, and then in Sec.\
\ref{sec3}, we present our results in studying SQM with this
model. Section \ref{sec4} is a short summary.

\begin{center}
\section{ Framework}
\end{center}

Let us schematically write the QCD Hamiltonian density as
\begin{equation}     \label{Hqcd}
  H_{\text{QCD}}=H_k + \sum_{q}m_{q0}\bar{q}q + H_{\text{I}},
\end{equation}
where $H_k$ is the kinetic term,
$m_{q0}$ is the quark current mass,
and $H_{\text{I}}$ is the interaction part. The summation goes over all
flavors considered.

The basic idea of the quark mass-density-dependent model of strange quark
matter is that the system energy can be expressed as the same form with
a proper noninteracting system. The strong interaction between quarks
is included within the appropriate variation of quark masses with density.
In order not to confuse with other mass concepts, we refer such a
density-dependent mass to an equivalent mass in this paper.
Therefore, if we use the equivalent mass $m_q$, the system Hamiltonian
density should be replaced by an Hamiltonian density of the form
\begin{equation}    \label{Hequiv}
  H_{\text{eqv}}=H_k + \sum_{q}m_{q}\bar{q}q,
\end{equation}
 where $m_q$ is the equivalent mass to be determined.
Obviously, we must require that the two Hamiltonian densities
$H_{\text{eqv}}$ and $H_{\text{QCD}}$ have the same eigenenergy
for any eigenstate $|\Psi\rangle$, i.e.,
\begin{equation}
  \langle{\Psi} | H_{\text{eqv}} |\Psi\rangle
 =\langle{\Psi} | H_{\text{QCD}} |\Psi\rangle.
\end{equation}
Applying this equality respectively to the state $|n_b\rangle$\
with baryon number density $n_b$ and the vacuum state $|0\rangle$,
and then taking the difference, one has
\begin{equation}
  \langle{n_b} | H_{\text{eqv}} | n_b\rangle
 -\langle{0} | H_{\text{eqv}} |0\rangle
 =\langle{n_b} | H_{\text{QCD}} | n_b\rangle
 -\langle{0} | H_{\text{QCD}} |0\rangle.
\end{equation}
%%where $|n_b\rangle$\ is the state vector with baryon number density $n_b$.
The simplest and most symmetric solution for the equivalent mass
from this equation is
\begin{eqnarray}
 m_q & = & m_{q0}+\frac{{\langle H_{\text{I}} \rangle}_{n_b}
                        -{\langle H_{\text{I}} \rangle}_{0}
                       }
                 {\sum\limits_{q}\left[{\langle\bar{q}q\rangle}_{n_b}
                                  -{\langle\bar{q}q\rangle}_{0}\right]
                 }
                                 \label{mqdef} \\
   & \equiv & m_{q0} + m_{\text{I}},          \label{midef}
\end{eqnarray}
where we have used the symbol definitions:
${\langle{H_{\text{I}}}\rangle}_{n_b}
\equiv \langle{n_b}|H_{\text{I}}|n_b\rangle$,
${\langle H_{\text{I}}\rangle}_{0}
\equiv \langle{0}|H_{\text{I}}|0\rangle$,
and
${\langle\bar{q}q\rangle}_{n_b}
\equiv \langle{n_b}|\bar{q}q|n_b\rangle$,
${\langle\bar{q}q\rangle}_{0}
\equiv \langle{0}|\bar{q}q|0\rangle$.

Therefore, if quarks are decoupled, they should take the
equivalent mass of the form (\ref{mqdef}) to keep the system
energy unchanged. From Eq.\ (\ref{mqdef}) we see that the
equivalent mass $m_q$ includes two parts: one is the original mass
or current mass $m_{q0}$, the other is the interacting part
$m_{\text{I}}$. Because $m_{\text{I}}$ equals to the ratio of the
total interacting part of the energy density and the total
relative quark condensate, it is flavor-independent and
density-dependent. Because of the quark confinement and the
asymptotic freedom, i.e.,
\begin{eqnarray}
  \lim_{n_b\rightarrow 0} m_{\text{I}} = \infty,  \\
  \lim_{n_b\rightarrow \infty} m_{\text{I}} = 0,
\end{eqnarray}
the reasonable form might be
\begin{equation}
  m_{\text{I}}=\frac{D}{n_b^z}.
\end{equation}
Accordingly, we have
\begin{equation}    \label{mq}
  m_q= m_{q0}+\frac{D}{n_b^z},
\end{equation}
where $D$ is a free parameter to be determined by stability
arguments. Obviously, $z>0$ for confined particles and $z<0$ for
nonconfined particles. In Eqs.\ (\ref{mudcha}) and (\ref{mscha}),
$z=1$. However, just as mentioned in the Introduction section, Eqs.\
(\ref{mudcha}) and (\ref{mscha}) is closely linked to the first
thermodynamic treatment, and thus unsuitable for our case.
%(Strictly speaking, these formulas should not be used for the second
%treatment as in Refs.\ \cite{Ben}).
We now discuss the determination of $m_{\text{I}}$ which is
consistent with our thermodynamic treatment.

Firstly, we express the interacting part of the energy density
$
\langle{H_{\text{I}}}\rangle \equiv
\langle{H_{\text{I}}}\rangle_{n_b} - \langle{H_{\text{I}}}\rangle_0
$
[the numerator in Eq.\ (\ref{mqdef})] as
\begin{eqnarray}
\langle{H_{\text{I}}}\rangle & = &
 \frac{1}{2V} \int\!\int_V v(r)
   (3n_b d\stackrel{\rightarrow}{r_1})
   (3n_b d\stackrel{\rightarrow}{r_2})    \\
 &=&
 18\pi n_b^2 \int_0^R v(r) r^2 dr,   \label{Hiapp}
\end{eqnarray}
where $r=|\stackrel{\rightarrow}{r_1}-\stackrel{\rightarrow}{r_2}|$,
$v(r)$ is the quark-quark interaction, $R$ is the SQM radius,
$V=4/3\pi{R^3}$ is the volume. The extra factor $1/2$ is responsible
for double counting.

Because of the following obvious  equality:
 \begin{equation}  \label{rqcl}
     \lim_{n_b\rightarrow 0}
   \frac{\langle\bar{q}q\rangle_{n_b}}
        {\langle\bar{q}q\rangle_0}=1,       \label{limitqc}
 \end{equation}
 the Taylor series of the relative
 condensate at zero density has the following general form:
 \begin{equation}   \label{eqpen}
   \frac{\langle\bar{q}q\rangle_{n_b}}
        {\langle\bar{q}q\rangle_0}
   =1-\frac{n_b}{\rho^\prime_{q}} +
              \mbox{\small higher orders in $n_b$} + \cdots.
 \end{equation}

If taking it only to first order approximation,
%% the denominator in Eq.\ (\ref{mqdef})
we have
\begin{equation}
\sum_q  \left[
{\langle\bar{q}q\rangle}_{n_b}-{\langle\bar{q}q\rangle}_{0}
               \right]
= \sum_q \left[
      -\langle\bar{q}q\rangle_0/\rho^\prime_q
                  \right] n_b
\equiv A n_b.                 \label{qc1}
\end{equation}

Taking the ratio of Eqs.\ (\ref{Hiapp}) and (\ref{qc1}), we get
\begin{equation}
m_{\text{I}} = \frac{18\pi}{A} n_b \int_0^R v(r) r^2 dr.
\end{equation}

According to the lattice calculation \cite{lattice} and string model
investigation \cite{string},
the quark-quark interaction is proportional to the distance, i.e.,
$v(r) = \alpha r$. We thus have
\begin{equation}
 m_{\text{I}} = \frac{18\pi\alpha}{A} n_b \frac{R^4}{4}
             \propto  \frac{1}{n_b^{1/3}}.
\end{equation}
Therefore, we should take in Eq.\ (\ref{mq})  $z=1/3$, i.e.,
\begin{equation}
m_q = m_{q0} + \frac{D}{n_b^{1/3}},   \label{mq13}
\end{equation}
where $D$ is a parameter to be determined by stability arguments.

Because the Hamiltonian density $H_{\text{eqv}}$ has the same form
as that of a system of free particles with equivalent mass $m_q$,
the energy density of SQM can be expressed as
\begin{equation}
  E = \sum\limits_{i=u,d,s,e} \frac{g_i}{2\pi^2}
      \int^{p_{f,i}}_0 \sqrt{p^2+m_i^2}\ p^2 dp+B,   \label{Edef}
\end{equation}
where
\begin{equation}
   p_{f,i}=\left(\frac{6}{g_i}\pi^2n_i\right)^{1/3}
\end{equation}
is the corresponding Fermi momentum.

As usually done, we here assume the SQM to consist of $u$, $d$, and $s$
quarks, and electrons (neutrinos enter and leave the system freely).
The degeneracy factor $g_i$ is 6 for quarks and 2 for electrons. The
electron mass $m_e$ is equal to 0.511 MeV. In order to include the strong
interaction between quarks, the quark masses $m_q (q=u,d,s)$ should
be replaced with the expression (\ref{mqdef}) or (\ref{mq13}). The extra
term $B$ comes from the pressure balance condition, and its physical meaning
is still the vacuum energy density or vacuum pressure
just as in the MIT bag model. The corresponding pressure is
\begin{equation}    \label{Ppen}
 P=\sum\limits_{i=u,d,s,e} \mu_i n_i - E,   \label{Pdef}
\end{equation}
where $\mu_i$ is the chemical potential for particle type $i$.
Because it is equal to the Fermi energy at zero temperature,
we have
\begin{equation}  \label{mupf}
  \mu_i=\sqrt{p_{f,i}^2+m_i^2}.
\end{equation}

Equation (\ref{Ppen}) is equivalent to
\begin{equation}
  P=-\Omega -B.
\end{equation}
The second term $-B$ is responsible for pressure balance. Such an
extra term is necessary even in the nonrelativistic treatment of
SQM \cite{Sat}.

 It is clear that the above thermodynamic treatment will approach the
conventional MIT bag model if one casts away the interacting part
$m_{\text{I}}$ of the equivalent mass $m_q$. It can be proved,
from Eqs.\ (\ref{mq13}), (\ref{Edef}), and (\ref{Pdef}), that we
have the following correct vacuum limits:
\begin{eqnarray}
  \lim\limits_{n_b\rightarrow{0}} E = B, \label{limbagE} \\
  \lim\limits_{n_b\rightarrow{0}} P = -B. \label{limbagP}
\end{eqnarray}
Therefore, the physical meaning of $B$ is the same as that in the
conventional bag model. We take $B^{1/4}=144$ MeV in our present
calculation.

\begin{center}
\section{Properties of strange quark matter } \label{sec3}
\end{center}

Following previous authors \cite{Jaffe}, we assume the SQM
to be a Fermi gas mixture of $u$, $d$, $s$ quarks and
electrons with chemical equilibrium maintained by the weak interactions:
$
 d, s \leftrightarrow u+e+\overline{\nu}_{e}, \, \,
                     s+u \leftrightarrow u+d.
$
For a given baryon number density $n_b$ and total electric charge density
$Q$, the chemical potentials $\mu_u$, $\mu_d$, $\mu_s$, and $\mu_e$
are determined by the following equations \cite{PengPRC59}:
\begin{eqnarray}
   &  \mu_d  = \mu_s \equiv \mu,  &    \label{eqmu1}     \\
   & \mu_u + \mu_e = \mu,         &                      \\
   & \frac{1}{3} (n_u + n_d + n_s) = n_b, &   \label{eqmu3}    \\
   & \frac{2}{3}n_u-\frac{1}{3}n_d-\frac{1}{3}n_s-n_e = Q,
                                      \label{eqmu4}   &
\end{eqnarray}
where the particle number density $n_i$ is related to the corresponding
chemical potential $\mu_i$ by
\begin{equation}
  n_i=\frac{g_i}{6\pi^2}(\mu_i^2-m_i^2)^{3/2},
\end{equation}
which is derived from the relation
\begin{equation}
   n_i=-\frac{\partial\Omega_i}{\partial\mu_i},
\end{equation}
with
\begin{eqnarray}                   \label{Omegai}
 & \Omega_{i}= -\frac{g_{i}}{48\pi^{2}}
\biggl[\mu_{i}(\mu_{i}^{2}-m_{i}^{2})^{1/2}(2\mu_{i}^{2}-5m_{i}^{2})&
                                              \nonumber   \\
  &+3m_{i}^{4}\ln\frac{\mu_{i}
         +\sqrt{\mu_{i}^{2}-m_{i}^{2}}}{m_{i}}\biggr].&
 \end{eqnarray}

In order to include the strong interaction between quarks, the quark
masses $m_u, m_d$, and $m_s$ in the above equations are to be replaced
with the density-dependent expression Eq.\ (\ref{mq13}) while the
electron mass $m_e$ is negligible (0.511 MeV).

For the bulk SQM with weak equilibrium, the previous
investigations got a slightly positive charge. Our recent study
demonstrates that negative charges could lower the critical
density. However, too much negative charge can make it impossible
to maintain flavor equilibrium. Therefore, the charge of SQM is
not allowed to shift too far away from zero at both positive and
negative directions. For this and our methodological purpose, we
only consider neutral SQM in this paper, i.e., $Q=0$ in Eq.\
(\ref{eqmu4}).

Since the baryonic matter is known to exist in the hadronic phase,
 we must require $D$ to be such that the $ud$ system is unbound.
This constrains $D$ to be bigger than (47 MeV)$^2$,
 i.e., at $P=0$, $E/n_b>930$ in order not to contradict standard
 nuclear physics. On the other hand, we are interested in the
 possibility that SQM might be absolutely stable, i.e., at $P=0$,
$E/n_b<930$, which gives an upper bound (128 MeV)$^2$. we take
$D^{1/2}$ to be 50, 80, and 110 MeV, respectively.

Because the light quark current masses are very small, their value
uncertainties are not important. So we take the fixed central
values $m_{u0}=5$ MeV and $m_{d0}=10$ MeV in our calculation. As for $s$
quarks, we take 150, 120, and 90 MeV, corresponding respectively
to $D^{1/2}$ = 50, 80, and 110 MeV.

For a given $n_b$, we first solve for $\mu_i\ (i=u,d,s,e)$ from
the equation group (\ref{eqmu1})--(\ref{eqmu4}), and then calculate
the energy density and pressure of SQM from Eqs.\ (\ref{Edef}) and
(\ref{Pdef}):
\begin{eqnarray}
 & E=\sum\limits_{i} \frac{g_i m_i^4x_i^3}{6\pi^2}F(x_i)+B
     =\sum\limits_{i} m_i n_i F(x_i)+B,      & \label{EchaP}   \\
 & P=\sum\limits_{i} \frac{g_i m_i^4x_i^5}{6\pi^2}G(x_i)-B
    =\sum\limits_{i} m_i n_i x_i^2 G(x_i)-B, &  \label{PchaP}
\end{eqnarray}
where the summation goes over $u, d, s$, and $e$, and
\begin{equation}
 x_i \equiv \frac{p_{f,i}}{m_i}=\frac{\sqrt{\mu_i^2-m_i^2}}{m_i}
\end{equation}
is the ratio of the Fermi momentum to the mass that related to
particle type $i$. With the hyperbolic sine function
sh$^{-1}(x)\equiv\ln(x+\sqrt{x^2+1})$, the functions $F(x)$ and
$G(x)$ are defined as
\begin{eqnarray}
 & F(x) \equiv \frac{3}{8}
    \left[
          x \sqrt{x^2+1}(2x^2+1)-\mbox{sh}^{-1}(x)
    \right]/x^3,  &   \\
 & G(x) \equiv \frac{1}{8}
    \left[
          x\sqrt{x^2+1}(2x^2-3)+3\mbox{sh}^{-1}(x)
    \right]/x^5,  &
\end{eqnarray}
which have the limit properties
\begin{eqnarray}
 & \lim\limits_{x\rightarrow 0} F(x) = 1, &   \\
 & \lim\limits_{x\rightarrow 0} G(x) = \frac{1}{5}.&
\end{eqnarray}
Therefore, we have the correct limits
Eqs.\ (\ref{limbagE}) and (\ref{limbagP}).

In Fig.\ \ref{enb}, we give the energy per baryon vs baryon number
density for the three pairs of parameters. We see that SQM is absolutely
stable for the first two parameter groups, while metastable for the
third group. The points marked with a circle ``$\bigcirc$'' are the zero
pressure points where the pressure within SQM is zero. Because of
the density dependence of quark masses, the zero pressure density is
generally not that corresponding to the minimum energy per baryon
(as in the usual case), but nearly the case in the first two
parameter groups.

    The resulting equation of state is plotted in Fig.\ \ref{eos}.
 Because it is insensitive to parameters, we have only chosen one
 parameter pair: $D=(80$ MeV)$^2$ and $m_{s0}=120$ MeV.

In Fig.\ \ref{soundv}, we show the sound velocity $c$ of SQM  with
a dot-dashed line, which is obtained from
\begin{equation}
  c=\left|\frac{dP}{dE}\right|^{1/2}.
\end{equation}
Because the interacting part of the quark masses is negligible at
higher densities, it asymptotically tends to the
ultrarelativistic value $1/\sqrt{3}$ as in the bag model (solid
line). Simultaneously given is that calculated by the same method
as in Ref.\ \cite{Ben} with $C=90$ MeV fm$^{-3}$ and $m_{s0}=80$
MeV (dotted line). Obviously, the lower density behavior of the
sound velocity in our model is opposite to that in the previous
calculation.

It is interesting to note that if one considers the thermodynamic
relation $P=-\partial(\Omega{V})/\partial{V}$ as being more
fundamental than $P=-\Omega$ (as done in Ref.\ \cite{Ben}), Eqs.
(\ref{EchaP}) and (\ref{PchaP}) should be replaced with
\begin{eqnarray}
 & E =\sum\limits_{i} m_i n_i F(x_i)
      +\sum\limits_{i} m_i n_i f(x_i) + B,      &   \\
 & P =\sum\limits_{i} m_i n_i x_i^2 G(x_i)
      -\sum\limits_{i} m_i n_i f(x_i) - B, &
\end{eqnarray}
where
\begin{equation}
 f(x_i) \equiv -\frac{3}{2}\frac{n_b}{m_i}\frac{dm_i}{dn_b}
                  \left[
               x_i \sqrt{x_i^2+1}-\mbox{sh}^{-1}(x_i)
                  \right]/x_i^3,
\end{equation}
which has the limit property
\begin{equation}
  \lim\limits_{n_b\rightarrow{0}} f(x_i) = z.
\end{equation}

However, the modification does not change the properties of SQM
significantly this time. For the same parameters, the line in
Fig.\ 1 will move upward slightly while in Fig.\ 2 and Fig.\ 3 a
little downward. This is because the contribution from the extra
term $\sum_{i} m_i n_i f(x_i)$ (arising from the density
dependence of the quark masses) is positive to energy and negative
to pressure. But the gross features of SQM are still the same.

\begin{center}
\section{ Summary } \label{sec4}
\end{center}

We have presented a new version of the quark
mass-density-dependent model for SQM. We first note that the
previous treatments have unreasonable vacuum limits. Then we
provide a practical  method to derive the quark mass formulas. In
our thermodynamic treatment, the conventional bag model is
included as an extreme, and the vacuum still has a constant energy
density corresponding to a constant pressure $B$.  In this new
treatment, SQM also has the possibility of absolute stability for
a wide range of parameters. A noticeable feature is that the sound
velocity is smaller than the ultrarelativistic case at lower
densities, contrary to the previous finding.

\begin{center}
\section*{          ACKNOWLEDGMENTS }
\end{center}

The authors would like to thank the National Natural Science Foundation
of China for partial financial support.

\begin{figure}
\caption{The energy per baryon vs baryon number density for
   different parameters. The zero pressure density occurs
   at the points marked with circle ``$\bigcirc$.''}
\label{enb}
\end{figure}

\begin{figure}
\caption{The equation of state for parameter group $D^{1/2}$ =
         80 MeV and $m_{s0}$ = 120 MeV.
         It asymptotically approaches to the
         ultrarelativistic case as expected. }
\label{eos}
\end{figure}
\begin{figure}
\caption{ The sound velocity vs energy density. The dot-dashed
          line is calculated with the method in this paper, while
          the dotted line is calculated with the same method in
          Ref.\ \protect\cite{Ben}. Their lower density behavior is
          obviously opposite. The full line is the
          ultrarelativistic case.
        }
\label{soundv}
\end{figure}

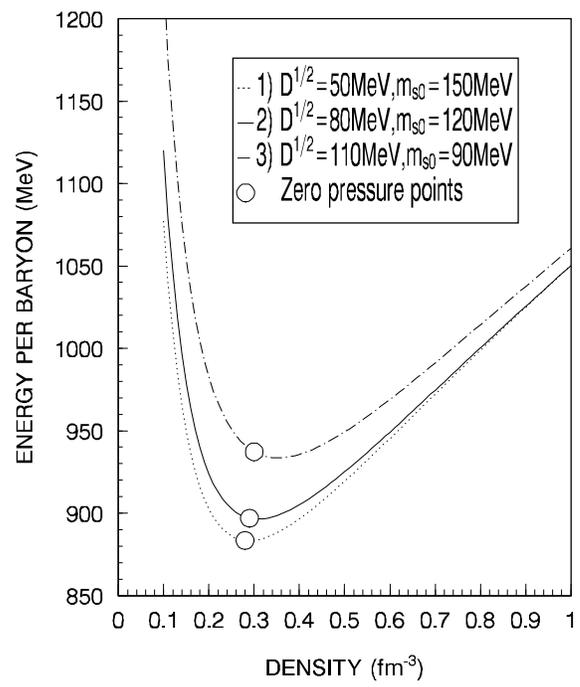
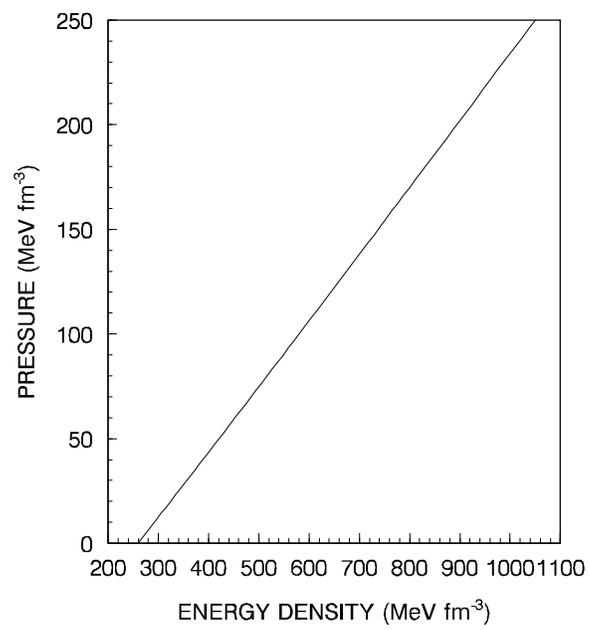
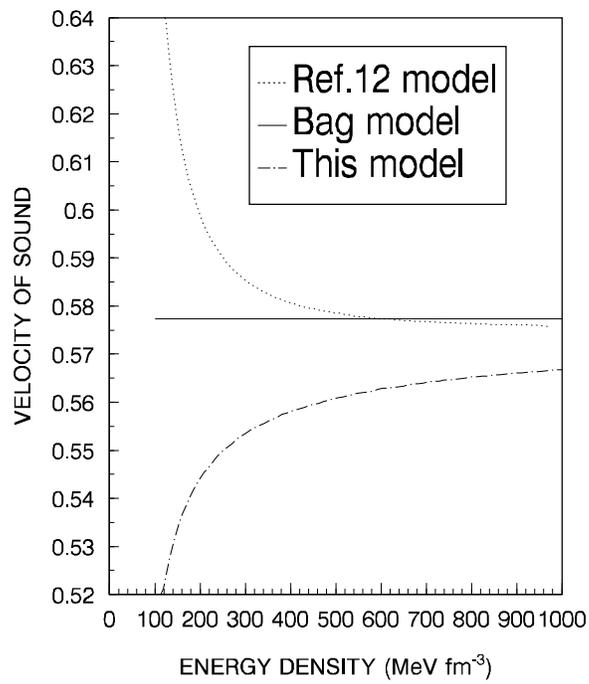
\end{document}